\documentclass[12pt,epsf,epsfig]{article}
\usepackage{amsmath}

\begin{document}

\begin{center}
{\huge{ Escaping the Hagedorn Temperature in Cosmology and Warped Spaces with Dynamical Tension Strings  }}  \\
\end{center}

\begin{center}
 E.I. Guendelman  \\
\end{center}

\begin{center}
\ Department of Physics, Ben-Gurion University of the Negev, Beer-Sheva, Israel \\
\end{center}

\begin{center}
\ Frankfurt Institute for Advanced Studies, Giersch Science Center, Campus Riedberg, Frankfurt am Main, Germany \\
\end{center}

\begin{center}
\ Bahamas Advanced Studies Institute and Conferences,  4A Ocean Heights, Hill View Circle, Stella Maris, Long Island, The Bahamas \\
\end{center}
E-mail:  guendel@bgu.ac.il,     

\abstract
In the modified measure formulation the string  tension appear as an additional dynamical degree of freedom and these tensions are not universal, but rather each string  generates its own  tension, which can have a different value for each string.  We consider new background field that can couple to these strings,  the ¨tension scalar¨ is capable of changing locally along the world sheet the value of the tension of the extended object. When many types of strings probing the same region of space are considered this tension scalar is constrained by the requirement of quantum conformal invariance. For the case of two types of strings probing the same region of space with different dynamically generated tensions, there are two different metrics, associated to the different strings,  that have to satisfy vacuum Einsteins equations and the consistency of these two Einstein´s equation detemines the tension scalar. The universal metric, common to both strings generically does not satisfy Einstein´s equation . 
In a previous paper we studied solutions that completely avoid singularities, but then one has to invoke positive and negative tension strings, which appear segregated in spacetime. In this paper we will consider only positive tension strings for the cosmological case and for warped space time . In both of these cases there is a region where the string tensions approach infinity, in the cosmological case this takes place in the early universe while in the warped case, for some value of the warping coordinate and since the Hagedorn temperature is proportional the the string tension, we get this way string scenarios with no limiting Hagedorn temperature in the early universe and this opens the possibility of a string cosmology without a Hgedorn phase transition through all of its history . Similar situation can take place in a warped space time, for this case, assuming the string have a tendency to avoid regions with lower Hagedorn temperature,  we obtain a mechanism for condensation of strings into a surface at high temperatures. 

\section{Introduction}

 String  Theories have been considered by many physicists for some time as the leading candidate for the theory everything,  including gravity, the explanation of all the known particles that we know and all of their known interactions (and probably more) \cite{stringtheory}. According to some, one unpleasant feature of string theory as usually formulated is that it has a dimension full parameter, in fact, its fundamental parameter , which is the tension of the string. This is when formulated the most familiar way.
The consideration of the string tension as a dynamical variable, using the modified measures formalism, which was  previously used for a certain class of modified gravity theories under the names of Two Measures Theories or Non Riemannian Measures Theories, see for example \cite{d,b, Hehl, GKatz, DE, MODDM, Cordero, Hidden}
Leads to the  modified measure approach to string theory, where  rather than to put the string tension by hand it appears dynamically.

This approach has been studied in various previous works  \cite{a,c,supermod, cnish, T1, T2, T3, cosmologyandwarped}. See also the treatment by Townsend and collaborators for dynamical string tension \cite{xx,xxx}.

In our most recent paper on the subject \cite{cosmologyandwarped},
we have also introduced the ¨tension scalar¨, which is an additional
background
fields that can be introduced into the theory for the bosonic case (and expected to be well defined for all types of superstrings as well) that changes the value of the tension of the extended object along its world sheet, we call this the tension scalar for obvious reasons. Before studying issues that are very special of this paper we review some of the material contained in previous papers,  first present the string theory with a modified measure and containing also gauge fields that like in the world sheet, the integration of the equation of motion of these gauge fields gives rise to a dynamically generated string tension, this string tensin may differ from one string to the other.

Then we consider the coupling of gauge fields in the string world sheet to currents in this world sheet, as a consequence this coupling induces variations of the tension along the world sheet of the string. Then we consider a bulk scalar and how this scalar naturally can induce this world sheet current that couples to the internal gauge fields. The integration of the equation of motion of the internal gauge field lead to the remarkably simple equation that the local value of the tension along the string is given by $T= e \phi + T _{i} $ , where $e$ is a coupling constant that defines the coupling of the bulk scalar to the world sheet gauge fields and  $ T _{i} $ is an integration constant which can be different for each string in the universe. 

Then each string is considered as an independent system that can be quantized. We take into account the string generation by introducing the tension as a function of the scalar field as a factor inside a Polyakov type action with such string tension, then the metric and the factor $g \phi + T _{i} $  enter together in this effective action, so if there was just one string the factor could be incorporated into the metric and the condition of world sheet conformal invariance will not say very much about the scalar  $\phi $ , but if many strings are probing the same regions of space time, then considering a background metric $g_{\mu \nu}$ , for each string the ¨string dependent metric¨  $(\phi + T _{i})g_{\mu \nu}$ appears and in the absence of othe background fields, like dilaton and antisymmetric tensor fields, Einstein´s equations apply for each of the metrics $(\phi + T _{i})g_{\mu \nu}$, considering two types of strings with $T _{1 \neq }T _{2}$. We call $g_{\mu \nu}$ the universal metric, which in fact does not necessarily satisfy Einstein´s equations.

 In the case of the flat space for the string associated metrics, in the Milne representation, for the case of two types of string tensions, we study the case where the two types of strings have positive string tensions , as opposed to our previous work \cite{cosmologyandwarped} where we found solutions with both positive and negative string tensions. At the early universe the negative string tension strings tensions are large in magnitude , but approach zero in the late universe and the  positive string tensions appear   for the late universe with their tension approaching a constant value at the late universe.  These solutions are absolutely singularity free.

In contrast, here we study the case of very different solutions and where both
type of strings have positive tensions,  then these are singular, they cannot be continued before a certain time (that corresponded to a bounce in our previous work \cite{cosmologyandwarped}). Here, at the origin of time, the string tensions of both types of strings approach plus infinity, so this opens the possibility of having no Hagedorn temperature \cite{Hagedorn} in the early universe and latter on in the history of the universe as well for this type of string cosmology scenario.    
 
The universal metric  has a singularity in the  cosmological case at the origin of time. 

Studying two types of strings  in warped space times of the type considered by Wesson and collaborators, leads to solutions where the string tensions approach to infinity as we approach a certain value of the wrapping coordinate. We argue that as a consequence at very high temperature strings try to concentrate close to this value of the wrapping coordinate where the Hagerdon temperature is very high, so as to escape the Hagedorn phase transition .

The notion of a Hagedorn temperature originates in a feature concerning the string model of hadrons, at there is a temperature 
where the partition function diverges \cite{HADRONHagedorn} and this was correctly interpreted as the breakdown of the string model of hadrons and that from temperatures above this Hagedorn temperature we should use QCD with quarks and gluons.

Still, as pointed out by Andreev \cite{Andreev}, the most simple string model of hadrons does not give the best results, since in the real world there is no phase transition but an analytic crossover, so he uses instead a multi tensions string model reproduces QCD  better, which is in fact an idea close in spirit to what we do here, we will comment more on this in our discussion section. 

For fundamental strings we do not have a more basic theory above the Hagedorn temperature, since the string theory itself is supposed  to be such fundamental theory. In the fundamental string theory the Hagedorn phase transition was studied for example by Atick and Witten and found to be first order, while other more recent a paper, by Brustein and Zigdon  contradicts Atick and Witten  and found the Hagedorn phase transition to be second order, but considers large expectation values for certain fields (that go like the inverse of the string coupling constant), that are nevertheless obtained in perturbation theory. See references to both papers among papers cited in   \cite{Hagedorn}. We do not want to enter this controversy, since our goal is to study a mechanism to in fact escape, this  Hagedorn temperature phenomenon.  

Whether this  Hagedorn temperature phase exists or not in some consistent string theories, we want to point out that this obviously makes string theory more complicated at high temperatures. This would be unfortunate, or at least, not along the trend in the progress of modern particle physics before string theory became dominant. Then we found for example the simplicity of QCD at high energy and the phenomenon of asymptotic freedom which makes the high energy phenomena much simpler to understand than the low energy hadronic physics.

Indeed from this experience the fundamental theory of nature should be simpler at high energies and higher temperatures, not more complicated, which means by all means avoiding the Hagedorn temperature. The dynamical tension string theory provide us indeed with this possibility as we will see in this paper.

This is naturally obtained by having the string tensions become very large in the early universe, which is the typical cosmological solution for this case, in the case the two types of string tensions are positive, since he string tension dictates the Hagedorn temperature, so if the string tension goes to infinity so does the Hagedorn temperature. Similar effect is obtained in warped space times,

The multi string tension model of Andreev \cite{Andreev}, mentioned before  that is used for describing Hadronic physics also avoids the Hagedorn temperature and certain aspects of Andreev´s approach will probably be incorporated into our type of multistring tension theories through the dynamical tension approach.\\

\section{The Modified Measure Theory String Theory}

The standard world sheet string sigma-model action using a world sheet metric is \cite{pol1}, \cite{pol2}, \cite{pol3}

\begin{equation}\label{eq:1}
S_{sigma-model} = -T\int d^2 \sigma \frac12 \sqrt{-\gamma} \gamma^{ab} \partial_a X^{\mu} \partial_b X^{\nu} g_{\mu \nu}.
\end{equation}

Here $\gamma^{ab}$ is the intrinsic Riemannian metric on the 2-dimensional string worldsheet and $\gamma = det(\gamma_{ab})$; $g_{\mu \nu}$ denotes the Riemannian metric on the embedding spacetime. $T$ is a string tension, a dimension full scale introduced into the theory by hand. \\

Now instead of using the measure $\sqrt{-\gamma}$ ,  on the 2-dimensional world-sheet, in the framework of this theory two additional worldsheet scalar fields $\varphi^i (i=1,2)$ are considered. A new measure density is introduced:

\begin{equation}
\Phi(\varphi) = \frac12 \epsilon_{ij}\epsilon^{ab} \partial_a \varphi^i \partial_b \varphi^j.
\end{equation}

There are no limitations on employing any other measure of integration different than $\sqrt{-\gamma}$. The only restriction is that it must be a density under arbitrary diffeomorphisms (reparametrizations) on the underlying spacetime manifold. The modified-measure theory is an example of such a theory. \\

Then the modified bosonic string action is (as formulated first in \cite{a} and latter discussed and generalized also in \cite{c})

\begin{equation} \label{eq:5}
S = -\int d^2 \sigma \Phi(\varphi)(\frac12 \gamma^{ab} \partial_a X^{\mu} \partial_b X^{\nu} g_{\mu\nu} - \frac{\epsilon^{ab}}{2\sqrt{-\gamma}}F_{ab}(A)),
\end{equation}

where $F_{ab}$ is the field-strength  of an auxiliary Abelian gauge field $A_a$: $F_{ab} = \partial_a A_b - \partial_b A_a$. \\

It is important to notice that the action (\ref{eq:5}) is invariant under conformal transformations of the intrinsic measure combined with a diffeomorphism of the measure fields, 

\begin{equation} \label{conformal}
\gamma_{ab} \rightarrow J\gamma_{ab}, 
\end{equation}

\begin{equation} \label{diffeo} 
\varphi^i \rightarrow \varphi^{'i}= \varphi^{'i}(\varphi^i)
\end{equation}
such that 
\begin{equation} \label{measure diffeo} 
\Phi \rightarrow \Phi^{'}= J \Phi
\end{equation}

Here $J$ is the jacobian of the diffeomorphim in the internal measure fields which can be an arbitrary function of the world sheet space time coordinates, so this can called indeed a local conformal symmetry.

To check that the new action is consistent with the sigma-model one, let us derive the equations of motion of the action (\ref{eq:5}). \\

The variation with respect to $\varphi^i$ leads to the following equations of motion:

\begin{equation} \label{eq:6}
\epsilon^{ab} \partial_b \varphi^i \partial_a (\gamma^{cd} \partial_c X^{\mu} \partial_d X^{\nu} g_{\mu\nu} - \frac{\epsilon^{cd}}{\sqrt{-\gamma}}F_{cd}) = 0.
\end{equation}

since $det(\epsilon^{ab} \partial_b \varphi^i )= \Phi$, assuming a non degenerate case ($\Phi \neq 0$), we obtain, 

\begin{equation} \label{eq:a}
\gamma^{cd} \partial_c X^{\mu} \partial_d X^{\nu} g_{\mu\nu} - \frac{\epsilon^{cd}}{\sqrt{-\gamma}}F_{cd} = M = const.
\end{equation}

The equations of motion with respect to $\gamma^{ab}$ are

\begin{equation} \label{eq:8}
T_{ab} = \partial_a X^{\mu} \partial_b X^{\nu} g_{\mu\nu} - \frac12 \gamma_{ab} \frac{\epsilon^{cd}}{\sqrt{-\gamma}}F_{cd}=0.
\end{equation}

One can see that these equations are the same as in the sigma-model formulation . Taking the trace of (\ref{eq:8}) we get that $M = 0$. By solving $\frac{\epsilon^{cd}}{\sqrt{-\gamma}}F_{cd}$ from (\ref{eq:a}) (with $M = 0$) we obtain the standard string eqs. \\

The emergence of the string tension is obtained by varying the action with respect to $A_a$:

\begin{equation}
\epsilon^{ab} \partial_b (\frac{\Phi(\varphi)}{\sqrt{-\gamma}}) = 0.
\end{equation}

Then by integrating and comparing it with the standard action it is seen that

\begin{equation}
\frac{\Phi(\varphi)}{\sqrt{-\gamma}} = T.
\end{equation}

That is how the string tension $T$ is derived as a world sheet constant of integration opposite to the standard equation (\ref{eq:1}) where the tension is put ad hoc.
Let us stress that the modified measure string theory action 
does not have any \textsl{ad hoc} fundamental scale parameters. associated with it. This can be generalized to incorporate super symmetry, see for example \cite{c}, \cite{cnish}, \cite{supermod} , \cite{T1}.
For other mechanisms for dynamical string tension generation from added string world sheet fields, see for example \cite{xx} and \cite{xxx}. However the fact that this string tension generation is a world sheet effect 
and not a universal uniform string tension generation effect for all strings has not been sufficiently emphasized before.

Notice that Each String  in its own world sheet determines its own  tension. Therefore the  tension is not universal for all strings.

\section{Introducing  Background Fields including a New Background Field, The Tension Field} 
Schwinger \cite{Schwinger} had an important insight and understood that all the information concerning a field theory can be studied by understanding how it  reacts to sources of different types. 

This has been discussed in the text book by Polchinski for example  \cite{Polchinski} .  Then the target space metric and other  external fields acquire dynamics which is enforced by the requirement of zero beta functions.

However, in addition to the traditional background fields usually considered in conventional string theory, one may consider as well an additional scalar field that induces currents in the string world sheet and since the current couples to the world sheet gauge fields, this produces a dynamical tension controlled by the external scalar field as shown at the classical level in \cite{Ansoldi}. In the next two subsections we will study how this comes about in two steps, first we introduce world sheet currents that couple to the internal gauge fields in Strings and Branes and second we define a coupling to an external scalar field by defining a world sheet currents that couple to the internal gauge fields in Strings  that is induced by such external scalar field.

\subsection{Introducing world sheet currents that couple to the internal gauge fields}

If to the action of the string  we add a coupling
to a world-sheet current $j ^{a}$,  i.e. a term
\begin{equation}
    S _{\mathrm{current}}
    =
    \int d ^{p+1} \sigma
        A _{a}
        j ^{a}
    ,
\label{eq:bracuract}
\end{equation}
 then the variation of the total action with respect to $A _{a }$
gives
\begin{equation}
    \epsilon ^{a b}
    \partial _{a }
    \left(
        \frac{\Phi}{\sqrt{- \gamma}}
    \right)
    =
    j ^{b}
    .
\label{eq:gauvarbracurmodtotact}
\end{equation}
We thus see indeed that, in this case, the dynamical character of the
brane is crucial here.
\subsection{How a world sheet current can naturally be induced by a bulk scalar field, the Tension Field}

Suppose that we have an external scalar field $\phi (x ^{\mu})$
defined in the bulk. From this field we can define the induced
conserved world-sheet current
\begin{equation}
    j ^{b}
    =
    e \partial _{\mu} \phi
    \frac{\partial X ^{\mu}}{\partial \sigma ^{a}}
    \epsilon ^{a b}
    \equiv
    e \partial _{a} \phi
    \epsilon ^{a b}
    ,
\label{eq:curfroscafie}
\end{equation}
where $e$ is some coupling constant. The interaction of this current with the world sheet gauge field  is also invariant under local gauge transformations in the world sheet of the gauge fields
 $A _{a} \rightarrow A _{a} + \partial_{a}\lambda $.

For this case,  (\ref{eq:gauvarbracurmodtotact}) can be integrated to obtain
\begin{equation}
  T =  \frac{\Phi}{\sqrt{- \gamma}}
    =
    e \phi + T _{i}
    ,
\label{eq:solgauvarbracurmodtotact2}
\end{equation}
or  equivalently
\begin{equation}
  \Phi
    =
   \sqrt{- \gamma}( e \phi + T _{i})
    ,
\label{eq:solgauvarbracurmodtotact}
\end{equation}

The constant of integration $T _{i}$ may vary from one string to the other. Notice tha the interaction is metric independent since the internal gauge field does not transform under the the conformal transformations. This interaction does not therefore spoil the world sheet conformal transformation invariance in the case the field $\phi$ does not transform under this transformation.  One may interpret 
(\ref{eq:solgauvarbracurmodtotact} ) as the result of integrating out classically (through integration of equations of motion) or quantum mechanically (by functional integration of the internal gauge field, respecting the boundary condition that characterizes the constant of integration  $T _{i}$ for a given string ). Then replacing 
$ \Phi
    =
   \sqrt{- \gamma}( e \phi + T _{i})$ back into the remaining terms in the action gives a correct effective action for each string. Each string is going to be quantized with each one having a different $ T _{i}$. The consequences of an independent quantization of  many strings with different $ T _{i}$
covering the same region of space time will be studied in the next section.    

\subsection{Consequences from World Sheet  Quantum Conformal Invariance on the Tension field, when several strings share the same region of space}

\subsubsection{The case where all all string tensions are the same, i.e.,  $T _{i}=  T _{0}$, and the appearance of a target space conformal invariance }
If  all  $T _{i}=  T _{0}$, we just redefine our background field so that $e\phi+T _{0} \rightarrow  e\phi$ and then in the effective action for all the strings the same combination $e\phi g_{\mu \nu}$, 
and only this combination will be determined by the requirement that the conformal invariance in the world sheet of all strings be preserved quantum mechanically, that is , that the beta function be zero. So in this case we will not be able to determine $e\phi$ and 
$ g_{\mu \nu}$ separately, just the product $e\phi g_{\mu \nu}$,
so the equation obtained from equating the beta function to zero will have the target space conformal invariance 
$e\phi \rightarrow F(x)e\phi $, 
$g_{\mu \nu} \rightarrow F(x)^{-1}g_{\mu \nu} $. 

That is, there is no independent dynamics for the Tension Field in this case.
On the other hand, if there are at least two types of string tensions, that symmetry will not exist and there is the possibility of determining separately  $e\phi$ and 
$ g_{\mu \nu}$ as we will see in the next subsection.

\subsubsection{The case of two different string tensions }

If we have a scalar field coupled to a string or a brane in the way described in the sub section above, i.e. through the current induced by the scalar field in the extended object,  according to eq. 
(\ref{eq:solgauvarbracurmodtotact}), so we have two sources for the variability of the tension when going from one string to the other: one is the integration constant $T _{i}$ which varies from string to string and the other the local value of the scalar field, which produces also variations of the  tension even within the string or brane world sheet.

As we discussed in the previous section, we can incorporate the result of the tension as a function of scalar field $\phi$, given as $e\phi+T_i$, for a string with the constant of integration $T_i$ by defining the action that produces the correct 
equations of motion for such string, adding also other background fields, the anti symmetric  two index field $A_{\mu \nu}$ that couples to $\epsilon^{ab}\partial_a X^{\mu} \partial_b X^{\nu}$
and the dilaton field $\varphi $ that couples to the topological density $\sqrt{-\gamma} R$
\begin{equation}\label{variablestringtensioneffectiveacton}
S_{i} = -\int d^2 \sigma (e\phi+T_i)\frac12 \sqrt{-\gamma} \gamma^{ab} \partial_a X^{\mu} \partial_b X^{\nu} g_{\mu \nu} + \int d^2 \sigma A_{\mu \nu}\epsilon^{ab}\partial_a X^{\mu} \partial_b X^{\nu}+\int d^2 \sigma \sqrt{-\gamma}\varphi R .
\end{equation}
Notice that if we had just one string, or if all strings will have the same constant of integration $T_i = T_0$.

In any case, it is not our purpose here to do a full generic analysis of all possible background metrics, antisymmetric two index tensor field and dilaton fields, instead, we will take  cases where the dilaton field is a constant or zero, and the antisymmetric two index tensor field is pure gauge or zero, then the demand of conformal invariance for $D=26$ becomes the demand that all the metrics
\begin{equation}\label{tensiondependentmetrics}
g^i_{\mu \nu} =  (e\phi+T_i)g_{\mu \nu}
\end{equation}
will satisfy simultaneously the vacuum Einstein´s equations,
Notice that if we had just one string, or if all strings will have the same constant of integration $T_i = T_0$, then all the 
$g^i_{\mu \nu}$ metrics are the same and then 
(\ref{tensiondependentmetrics}) is just a single field redefinition and therefore there will be only one metric that will have to satisfy Einstein´s equations, which of course will not impose a constraint on the tension field $\phi$ . 

The interesting case to consider is therefore many strings with different $T_i$, let us consider the simplest case of two strings, labeled $1$ and $2$ with  $T_1 \neq  T_2$ , then we will have two Einstein´s equations, for $g^1_{\mu \nu} =  (e\phi+T_1)g_{\mu \nu}$ and for  $g^2_{\mu \nu} =  (e\phi+T_2)g_{\mu \nu}$, 

\begin{equation}\label{Einstein1}
R_{\mu \nu} (g^1_{\alpha \beta}) = 0 
\end{equation}
and , at the same time,
\begin{equation}\label{Einstein1}
  R_{\mu \nu} (g^2_{\alpha \beta}) = 0
\end{equation}

These two simultaneous conditions above  impose a constraint on the tension field
 $\phi$, because the metrics $g^1_{\alpha \beta}$ and $g^2_{\alpha \beta}$ are conformally related, but Einstein´s equations are not conformally invariant, so the condition that Einstein´s equations hold  for both  $g^1_{\alpha \beta}$ and $g^2_{\alpha \beta}$
is highly non trivial.

Let us consider the case that one of the metrics, say  $g^2_{\alpha \beta}$ is a Schwarzschild solution, either a 
4 D Schwarzschild solution X a product flat of Torus compactified extra dimensions
or just a 26 D Schwarzschild solution, in this case, it does not appear possible to have a conformally transformed  $g^2_{\alpha \beta}$ for anything else than in the case that the conformal factor that transforms the two metrics is a positive constant, let us call it $\Omega^2$, in that case $g^1_{\alpha \beta}$ is a Schwarzschild solution of the same type, just with a different mass parameter and different sizes of extra dimensions if the compactified solution is considered. Similar consideration holds for the case the 2 metric is a Kasner solution,

Then  in this case also, it does not appear possible to have a conformally transformed  $g^2_{\alpha \beta}$ for anything else than in the case that the conformal factor that transforms the two metrics is a constant, we will find other cases where the conformal factor will not be a constant,  let us call then conformal factor $\Omega^2$ in general, even when it is not a constant.
One can also study metrics used to describe gravitational radiation,
then again, multiplying by a constant both the background flat space and the perturbation gives us also a solution of vacuum Einstein´s equations. 

Then for these situations, we have,
\begin{equation}\label{relationbetweentensions}
e\phi+T_1 = \Omega^2(e\phi+T_2)
\end{equation}
 which leads to a solution for $e\phi$
 
\begin{equation}\label{solutionforphi}
e\phi  = \frac{\Omega^2T_2 -T_1}{1 - \Omega^2} 
\end{equation}
which leads to the tensions of the different strings to be
\begin{equation}\label{stringtension1}
 e\phi+T_1 = \frac{\Omega^2(T_2 -T_1)}{1 - \Omega^2} 
\end{equation}
and
  \begin{equation}\label{stringtension2}
 e\phi+T_2 = \frac{(T_2 -T_1)}{1 - \Omega^2} 
\end{equation}

Both tensions can be taken as positive if $T_2 -T_1$ is positive and $\Omega^2$ is also positive and less than $1$.
It is important that we were forced to consider a multi metric situation. One must also realize that the constant $c$ is physical, 
because both metrics live in the same spacetime, so even if c is a constant ,  we are not allowed to perform a coordinate transformation, consisting for example of a rescaling of coordinates for one of the metrics and not do the same transformation for the other metric. 

Other way  to see that $\Omega^2$ is physical consist of considering the scalar consisting of the ratio of the two measures $\sqrt{-g^1}$ and $\sqrt{-g^2}$ where $ g^1 = det ( g^1_{\alpha \beta})$ and $ g^2 = det ( g^2_{\alpha \beta})$, and we find that the scalar 
$\frac{\sqrt{-g^1}}{\sqrt{-g^2}} = \Omega^{D}$, showing that $\Omega$ is a coordinate invariant. 

Let us study now a case where $\Omega^2$ is not a constant, we will also focus on a cosmological case. To find this it is  useful to consider flat space in the Milne representation, $D=4$ this reads,
 \begin{equation}\label{Milne4D}
 ds^2 = -dt^2 + t^{2}(d\chi^2 + sinh^2\chi d\Omega_2^2)
\end{equation}

where $ d\Omega_2^2 $ represent the contribution of the 2 angles to the metric when using spherical coordinates, that is, it represents the metric of a two dimensional sphere of unit radius. In $D$ dimensions we will have a similar expression but now we must introduce the metric of a  $D-2$ unit sphere $ d\Omega_{D-2}^2 $ so we end up with the following metric that we will take as the metric 2
 \begin{equation}\label{MilneD1}
 ds_2^2 = -dt^2 + t^{2}(d\chi^2 + sinh^2\chi d\Omega_{D-2}^2)
\end{equation}

For the metric $1$ we will take the metric that we would obtain from the coordinate $t \rightarrow 1/t $ (using Minkowskii coordinates 
$x^\mu $, this corresponds to the inversion transformation, for a review and generalizations see  \cite{Kastrup}.
 $x^\mu \rightarrow x^\mu/(x^\nu x_\nu ) $) and then we furthermore multiply by a constant $\sigma$, so

 \begin{equation}\label{MilneD2}
 ds_1^2 =\frac{\sigma}{t^4} (-dt^2 + t^{2}(d\chi^2 + sinh^2\chi d\Omega_{D-2}^2))
\end{equation}

Then the equations (\ref{relationbetweentensions}), (\ref{solutionforphi}), (\ref{stringtension1}), (\ref{stringtension2}),
with $ \Omega^2= \frac{\sigma}{t^4}$. In \cite{cosmologyandwarped} to avoid possible singularities, we took $\sigma$ negative,
and therefore we had two strings tensions with opposite opposite signs. Here, by writing the the conformal factor between the metrics as  $ \Omega^2$, we are explicity avoiding negative conformal factors and furthermore, all string tensions are taken to be positive.

While avoidance of singularities is very interesting, there is the need to introduce negative tension strings. 

As we will see here considering same sign string tensions introduces cosmological singularities, but on the other hand we obtain at that singular point an infinite positive string tension for the two string tensions, which imply absence of a Hagedorn temperature in the early universe. There will not be any obstacle for the temperature of the universe to become infinte. We consider now $ \Omega^2== \frac{\sigma}{t^4}$,
with $\sigma$ and therefore $\Omega^2$ positive.

The strings 1 and 2 have both positive tensions if the sign of $T_2 -T_1$ is positive and the solution is not continued before the singularity.

\subsection{The  Singular Behavior for the Universal
Metric and Recovery of Target Space Conformal Invariance For the Early Universe}
As we have seen when the space time is probed
by two types of strings,
there are two metrics that have to satisfy the vacuum Einstein´s equations, this is enough to solve the problem, The interesting thing however is that the universal metric $ g_{\mu \nu}$
does  not have to satisfy Einstein´s equation. We can see this by solving  $ g_{\mu \nu}$ in terms of one of the metrics , for example from  $g^2_{\mu \nu} =  (e\phi+T_2)g_{\mu \nu}$, we have that

 \begin{equation}\label{universalmetric}
 ds^2 =g_{\mu \nu}dx^{\mu}dx^{\nu} =  (\frac{{1 - \Omega^2}}{T_2 -T_1})(-dt^2 + t^{2}(d\chi^2 + sinh^2\chi d\Omega_{D-2}^2))
\end{equation}
and considering that  $\Omega^2= \frac{\sigma}{t^4}$ ,
where $\sigma$ is positive. So the coefficient of the hyperbolic D -1 dimensional metric $d\chi^2 + sinh^2\chi d\Omega_{D-2}^2$ is 
$\frac{{t^{2} - \frac{\sigma}{t^2}}}{T_2 -T_1}$, showing a collapse at $t =t^*= (\sigma)^{1/4}$. 

We can expand  the scale factor around $t =t^*= (\sigma)^{1/4}$,
defining  $t=t^*+ \bar{t}$.
The result is, just keeping the first linear term in  $ \bar{t}$, 
\begin{equation}\label{universalmetricEarlyU}
\frac{{t^{2} - \frac{\sigma}{t^2}}}{T_2 -T_1} = \frac{4(\sigma)^{1/4}}{T_2 -T_1}\bar{t}
\end{equation}
One can bring this metric in terms of cosmic time, where the $00$ of  metric component of $g_{\mu \nu}$ is normalized to $1$. One has to notice however that we cannot do the transformation of coordinates only on one of the three metrics we have discussed $g^2_{\mu \nu}, g^1_{\mu \nu}$ and $g_{\mu \nu}$, and if we bring the  the $00$ metric component of the metric $g_{\mu \nu}$ is normalized to $1$, it will not happen simultaneously for  $g^2_{\mu \nu}, g^1_{\mu \nu}$. Having this in mind, the cosmic time coordinate $T$ where  where the $00$ of  metric component of $g_{\mu \nu}$ is normalized to $1$ is defined by 
 \begin{equation}\label{cosmictime}
 dT  = \sqrt{ \frac{{1 - \frac{\sigma}{t^4}}}{T_1 -T_2}}dt 
\end{equation}
and close to the singularity of the universal metric, $t =t*= (\sigma)^{1/4}$, we see tat theb relation between, $T$ and  $ \bar{t}$
is  $ \bar{t} \rightarrow  A (T)^{2/3}$, where $A$ is a constant.
So, we see that is the dependence of the scale factor close to the singularity.
Since there is a singularity , not only for the universal metric, but also for the tensions of the two types of strings, we do not extend the solution for times before the singularity. If we did that, also the tensions of the two types of strings become negative.

Notice that for the Early Universe, near the point where the tensions of both strings diverge, the ratio between the tensions of both strings become one, so this appears as a restoration of the target space conformal invariance discussed before, valid for the case where the string tensions are equal.
\subsection{ Absence of a Hagedorn Temperature at Early times in the cosmological case, vanishing slope parameters in the early universe as a sign of asymptotic freedom}
We notice that at the singular point, $t =t^*= (\sigma)^{1/4}$, where  $\Omega^2 \rightarrow  1 $, and from the expressions of the tensions of strings 1 and 2, ( eqs. \ref{stringtension1}, \ref{stringtension2}), that both string tensions become arbitrarily large at this point. Since the Hagedorn Temperature is proportional to the string tension, we conclude that in the early Universe the will no maximum temperature or Hagedorn phase transition. 
At the early universe, the slopes,
 $\alpha_1' = 1/4\pi T_1 $ and  $\alpha_2' = 1/4\pi T_2 $ are very small, so the expansion that gives the effective gravity equatinsfrom the requirement that  the conformal invariance is preserved at the quantum level is very reliable, since this relies in a perturbative expansion in the slopes,
 $\alpha_1' = 1/4\pi T_1 $ and  $\alpha_2' = 1/4\pi T_2 $.
 
 We  can say therefore that there is a kind of asymptotic freedom of this theory for the early universe, which is exactly the reason that we are relieved (or allowed to escape) from the Hagerdorn temperature  
 in the early universe.

\subsection{The Late Universe Cosmology}
For the late universe, there is no thermal equilibrium and there are two types of strings, one type has a practically constant string tension , the strings type 2  ( eq. \ref{stringtension2}) and the other type of string  , the strings type 1  ( eq. \ref{stringtension1}) have a very small string tension, suppressed by the factor of $\Omega^2 \rightarrow  0 $ as $ t \rightarrow  \infty  $ . 
One should investigate further whether there is a special role of these ultralight strings in the late universe in cosmology and particle physics.

As opposed to the early universe, one of the slopes, the one associated to the string 1 becomes very big, so the physics of that type of string becomes probably non perturbative at this very late times (and low energies), since the effective gravity theory relies on an expasion on the slope, and now the slope of the light strings is rather big. At leas non perturbative appear now in the low enegy sector, the dynamical tension theory removes the non perturbative effects at high temperature and place now son non perturbative effects at low energies, that is ok with the general intuition which is the basis of this paper.

\subsection{ Spontaneously Generated Boundary in Wesson warped spaces}

One may wonder if there are similar solutions to the vacuum Einstein´s equations similar  to the Milne space but where instead of time some spacial coordinate would play a similar way. The answer to this question is yes, and these  are the solutions in higher dimensional vacuum General Relativity discovered by Wesson and collaborators, see
\cite{Wesson} and references there. In  five dimensions for example the following warped solution is found,
\begin{equation}\label{Wesson1}
 ds^2 =l^2dt^2 -l^{2} cosh^{2}t (\frac{dr^2}{1-r^{2}} 
 + r^2 d\Omega_{2}^2)) - dl^{2}
\end{equation}
where $l$ is the fourth dimension,
so we see that as in the fourth dimension  $l$  such a solution is homogeneous of degree two, just as the Milne space time was homogeneous of degree two with respect to the time. Notice that maximally symmetric de Sitter space times sub spaces $l =$ constant appear for  instead of euclidean spheres that appear in the Milne Universe for $t =$ constant.

The list of space times of this type is quite large, for example, one cal find solutions of empty GR with Schwarzchild de Sitter subpaces for  $l =$ constant, as in
\begin{equation}\label{Wesson2}
 ds^2 =\frac{\Lambda l^{2}}{3} (dt^2 (1-\frac{2M}{r} -\frac{\Lambda r^2}{3}) - \frac{dr^2}{1-\frac{2M}{r} 
 -\frac{\Lambda r^2}{3}} 
 - r^2 d\Omega_{2}^2) - dl^{2}
\end{equation}
This of course can be extended to $D$ dimensions, where we choose one dimension $l$ to have a factor $l^2$ warp factor for the other dimensions , generically for  $D$ dimensions as in 
\begin{equation}\label{Wessongeneric}
 ds_2^2 =l^{2}\bar{g}_{\mu \nu}(x)dx^{\mu}dx^{\nu} - dl^{2}
\end{equation}
where $\bar{g}_{\mu \nu}(x)$ is a $D-1$ Schwarzschild de Sitter metric for example \cite{Wesson}. This we will take as our $2$ metric, 

In any case, working with this generic  metric of the form  (\ref{Wessongeneric}), but now in $D$ dimensions,  we can perform the inversion transformation $l \rightarrow \frac{1}{l} $, and multiplying also by a factor $ \sigma$ and obtain the conformally transformed metric $1$ that also satisfies the vacuum Einstein´s equations
\begin{equation}\label{Wessongenericinverted}
 ds_1^2 = \sigma l^{-2}\bar{g}_{\mu \nu}(x)dx^{\mu}dx^{\nu} - \sigma\frac{dl^{2}}{l^{4}} = \sigma l^{-4}ds_2^2
\end{equation}

From this point on , the equations the solutions for the tensions of the $1$ and $2$ strings are the same as in the cosmological case, just that $t  \rightarrow l$, so now $\Omega^2=  \sigma l^{-4}$, so that we now insert this  expression for $c$ in (\ref{stringtension1}) and in  
(\ref{stringtension2}). 

Now , we will choose  $\sigma $ positive, since we work here only with two types of strings, both with positive tension. obtaining that on one value of the wrapping coordinate in $l$ both string tensions approach arbitrarily large values.

The universal metric, following the steps done for the cosmological case is now ,
\begin{equation}\label{universalmetricwithldependence}
 ds^2 =  (\frac{{1 - \Omega^2}}{T_2 -T_1}) (l^{2}\bar{g}_{\mu \nu}(x)dx^{\mu}dx^{\nu} - dl^{2})
\end{equation}

looking at the coefficient of $\bar{g}_{\mu \nu}(x)dx^{\mu}dx^{\nu}$, the function is $(\frac{{l^{2} - \sigma/l^{2}}}{T_2 -T_1}) $, so the space
time is expanded or contracted as we move in the dimension $l$ by this factor. This factor explodes at $l^* = \sigma^{1/4}$.
We can define a proper length  coordinate $L$ where  where the $ll$ of  metric component of the metric is normalized to $-1$ is defined by 
 \begin{equation}\label{Properlength}
 dL  = \sqrt{ \frac{{1 -\frac{\sigma}{l^4}}}{T_2 -T_1}}dl 
\end{equation}

So, we see that as $l \rightarrow + \infty  $, 
 $L \rightarrow c_1 l$, while for  $l \rightarrow l^* = \sigma^{1/4} $,  
 $L \rightarrow c_2(l- l^*)^{3/2}$, where $c_2$ is a constant, or equivalently, $(l- l^*)$ goes as $L^{2/3}$ and the coefficient of the $\bar{g}_{\mu \nu}(x)$ metric goes as $L^{4/3}$.
 
 The point $l=l^*$ is a spontaneously generated boundary, where a singularity occurs. In the next section we discuss the possible role of this surface in a condensation of strings at ultrahigh temperatures.
 
 \subsection{ Absence of a Hagedorn Temperature at the spontaneously generated boundary, condensation of strings at ultrahigh temperatures. }
 
The tensions of the two types of string tensions approach plus infinity as we let $l \rightarrow l^*$, since then $c\rightarrow 1 $.
As we discussed in the cosmological case, this implies that the Hagedorn temperature, which is proportional to the string tension also diverges at this value of the wrapping coordinate $l$.

If the universe is subjected to a very high temperature, it would be reasonable to expect that the strings would prefer to move to a region of space with no maximum Hagedorn temperature or phase transition, that is, they would condensate close to  $l = l^*$.
 
 \section{Conceptual issues to be resolved: is entanglement in action here and is this a quantum version of Mach principle? }
 
 We notice the quantum conformal invariance starts to give useful information concerning the tension field only after at least two strings with different string tensions covering the same region of space are considered. Then the behavior of one string appear to correlated with the other, it dictates for that othe s string what its tension should be, and vice versa. Since this correlation is achieved through quantum mechanics, it seems legitimate to say that this is a kind on entanglement.
 
 Furthermore, since what is being correlated are the tensions of different strings, this certainly resembles the notion that this is a version of Mach principle in action. Recall that Mach principle suggest that a mass in the universe is determined by all the other masses in the universe. In this case, we talk about the tension of the strings instead of masses.  
 
 \section{ Possible Applications for the construction  of a Hagedorn Temperature free String Model of the Strong interactions. }

 In \cite{Andreev}, Andreev has discussed the need to avoid the Hagedorn temperature in order to obtain a behavior more in accordance to that of QCD, since in the real world there is no phase transition but an analytic crossover. If strings are indeed relevant for QCD then one has to show that a stringy
description is also valid for high T and he has shown that this can still be achieved but in the context of string models, but then these string models have to be multy tension string models, so it appear that our ideas and those of  Andreev go in the same direction.

The approach by Andreev  is more phenomenological than ours and he introduces strings with different tensions and then gives a prescription concerning the string tensions that one should be allowed to contribute at a given temperature. Such type of prescriptions will probably not be necessary in our approach, no need to eliminate states with certain string tensions depending of the temperature,  instead we would rely on the the  dynamical effect where the tensions grow and suppress the Hagedorn temperature. Much work in this direction is needed.

 \section{Discussion, comparing the singularity free but with negative tensions or a singular but with an escape from the Hagedorn Temperature option and only positive tensions scenarios? }
 
The main results here in this paper concern the behavior of the tension scalar and the consequent behavior of the string tensions  for two types of strings with two different string tensions which are both positive. The results differ from our previous work where the case of two types of strings with opposite signs were introduced \cite{cosmologyandwarped}. The choice of introducing negative tension strings was with the goal of avoiding singularities both in the cosmological case as well as in the wrapped space times. That goal was achieved indeed.

Here we insist in scenarios where both type of strings have positive tension, as a result a singularity appears in the behavior of the tensions. At some point in time, in the cosmological case, or at some value of the wrapping coordinate in the warp space scenario. As a consequence , at high temperatures, strings can always avoid the Hagedorn maximal temperature, or phase transition,  at least in a part of the space time.

Avoidance of singularities and has been studied in the context of the Einstein´s equations, where a cosmological singularity generically applies when positive energy conditions are satisfied, but can be avoided when introducing exotic matter. Here, although Einstein´s equations are not directly applicable, since the universal metric does not satisfy Einstein´s equations, a singularity is avoided also by introducing negative tension strings in the early universe \cite{cosmologyandwarped}, similar situation holds in the warped scenario, which is also devoid of singularities. So the advantages of the scenario explained in \cite{cosmologyandwarped} are clear: avoidance of singularities at the price of introducing negative tension strings.

The advantages of the scenario presented in this paper in comparison are: first,  no negative tension strings are invoked, second, in the region where both tensions are very big, the calculation of the constraint that the beta function is zero, which a perturbative calculation in the inverse of the tension, that is the slopes,
 $\alpha_1' = 1/4\pi T_1 $ and  $\alpha_2' = 1/4\pi T_2 $ are very small, so the expansion is very reliable, third: we obtain  a remarkable new effect, no Hagedorn limiting temperature in the early universe in cosmology or close to a certain value of the wrapping coordinate in the wrapped space time scenario, fourth: in the wrapped scenario, at ultra high temperatures, the strings presumably tend to to be in the Hagedorn free region of space, i.e., located very close to the value of the wrapping coordinate where the string tensions blow up, producing a condensation of the strings into 
the surface $l=l^*$.

Of course much work remains to be done, some of which was discussed in 
 \cite{cosmologyandwarped}  for the scenario with negative tension strings, but also concerning general features of the approach that also apply here. For the wrapped scenario an important aspect concerns the production of gravity waves, as studied in \cite{warpgravitywaves}
. A corresponding study or our models has to be performed as well.
 
\textbf{Acknowledgments}
 I thank Oleg Andreev, David Andriot, Stefano Ansoldi, Euro Spallucci, Emil Nissimov, Svetlana Pacheva, Tatiana Vulfs,  Hitoshi Nishino, Subhash Rajpoot and David Benisty for usefull discussions. I also want to thank the Foundational Questions Institute (FQXi)  and the COST actions  Quantum Gravity Phenomenology in the multi messenger approach, CA18108 and  Gravitational waves, Black Holes and Fundamental Physics, CA16104 for support.

\end{document}